\documentclass{article}

\usepackage{multicol}
\usepackage{geometry}

\usepackage{hyperref} 
\hypersetup{
    colorlinks=true,
    linkcolor=blue,
    filecolor=black,      
    urlcolor=black,
    pdftitle={Overleaf Example},
    pdfpagemode=FullScreen,
    citecolor = blue,
    }
\usepackage{graphicx} 
\usepackage{amsmath}
\usepackage{caption}
\usepackage{url}  

\title{Solving Free Boundary Problems in Alloy Solidification \\ under Universal Cooling Conditions}

\author{Chuanqi Zhu\footnote{ Email: \url{chuanqizhu820@outlook.com}}
\and Yuichiro Koizumi}

\date{February 2024}

\begin{document}

\maketitle

\begin{center}
\section*{Abstract}
\end{center}
The kinetics of interfaces in alloy solidification pose a classic free boundary problem. This paper introduces an approach that amalgamates the distinctive characteristics of sharp and diffuse interface models. The motion of the diffuse interface is governed by the phase-field equation featuring a traveling wave function [I. Steinbach, Modell. Simul. Mater. Sci. Eng. \textbf{17}(7), 073001 (2009)]. To emulate solute rejection in the sharp interface model, the concept of the ``middle obstacle'' and the casting operation are employed. Moreover, the undercooling along the interface normal is flattened to minimize the impact of bulk undercooling in the interface and the associated effects, such as stretching and arc-length diffusion. Notably, the results for interface kinetics under both equilibrium and non-equilibrium conditions closely approximate their analytical solutions, all achieved with artificially wide interfaces and comparatively low computational costs\footnote{Access to the sample code: https://github.com/chuanqizhu820/FBP-ALLOY}. 
\newline

\section{Introduction}
The free boundary problem is formulated by the partial differential equation for indeterminate domains with a moving boundary. In the field of solidification, it is known as the sharp interface model that describes the motion of the solid-liquid interface during the phase transition accompanied by heat or mass transport \cite{rappaz2009solidification}. The front-tracking \cite{juric1996front, kim2000computation,zhu2007virtual, tan2007level, wei2019quantitative} and phase-field \cite{kobayashi1994numerical, provatas1998efficient, takaki2014phase, karma2016atomistic} methods are two distinct approaches for solving these problems in the solidification of industrial alloys. In the regime of the slow cooling rate, the phase-field method has gained its success \cite{steinbach2009phase, steinbach2013solidification, tourret2022phase} owing to the establishment of quantitative methodology \cite{karma2001phase, ohno2009quantitative}. However, when the cooling rate becomes high, the accurate incorporation of non-equilibrium effects, such as solute trapping and solute drag \cite{aziz1982model, aziz1994transition, yang2011atomistic} has long been a challenge, which has promoted the development of a variety of phase-field models for solving non-equilibrium interface kinetics \cite{steinbach2012phase,pinomaa2019quantitative, kavousi2021quantitative, ji2023microstructural,huang2023comparative, mukherjee2023quantitative}. Faced with the industrial challenges in the next decades \cite{wong2012review, koizumi2022digital}, there is a clear demand for advanced modeling techniques of alloy solidification that can cover a wide range of cooling conditions.

Rather than extending previous phase-field models, the present work implemented an idea that combines the salient features of both front-tracking and phase-field models \cite{zhu2023middle}. Surprisingly, the numerical solutions of the interface kinetics under both equilibrium and non-equilibrium conditions can approximate the true solutions of the sharp interface models. The four principles of the proposed model are: (1) The equation for the motion of the diffuse interface is formulated in alignment with the phase-field method; (2) Fick's diffusion equations in the solid and liquid phases are solved separately by assuming the interface is a mixture of the two phases; (3) The concentrations in the two phases are related by the equilibrium or non-equilibrium partition coefficient; (4) The process of solute rejection in the sharp interface model has been emulated through the cast operation on the rejected solute behind the so-called ``middle obstacle''. The following content aims to clarify the simple and effective methodology for solving free boundary problems in alloy solidification under universal cooling conditions.

\section{Methods}

\subsection{Standard alloy model}

\begin{figure}[htbp]
    \centering
    \includegraphics[width=0.7\linewidth]{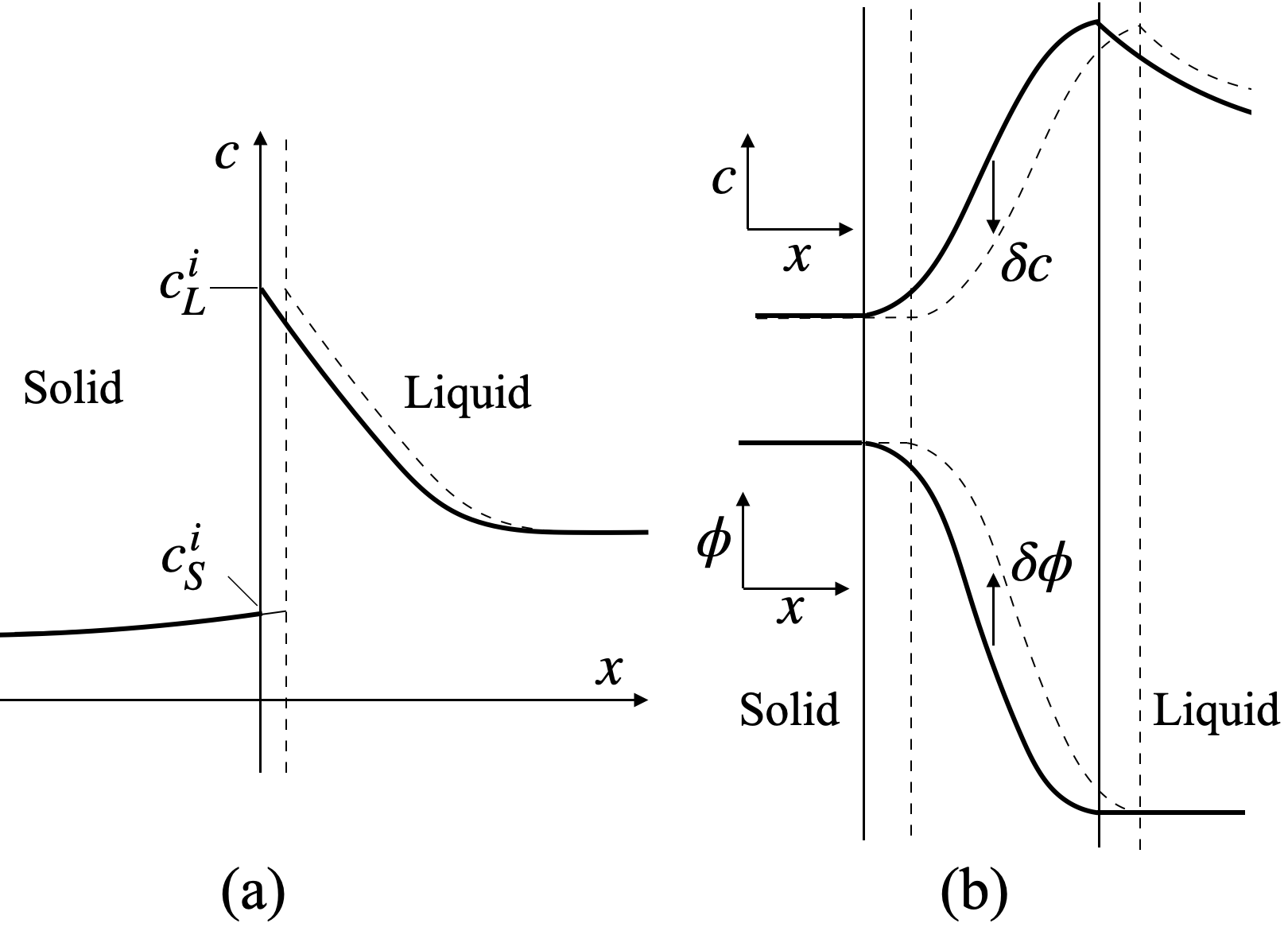}
    \caption{ Schematics of sharp and diffuse interface models for alloy solidification. (a) Concentration profile across the sharp interface; (b) coupling of phase and concentration fields in the diffuse interface model. The dashed lines denote the shifted profiles of the moving interfaces.}
    \label{fig:models}
\end{figure}

The phase-field equation with traveling wave function is given as 
\begin{equation}
    \frac{\partial \phi}{\partial t} =  \mu \bigg \{ \Gamma \bigg[ \nabla^2\phi +\frac{\pi^2}{\eta^2} \bigg(\phi-\frac{1}{2}\bigg)\bigg]+ \frac{\pi\sqrt{\phi(1-\phi)}}{\eta}\Delta T \bigg \}, 
    \label{eq:dphidt}
\end{equation}

\noindent which can be equivalent to the interface kinetic equation with Gibbs-Thomson effect \cite{steinbach2009phase}. Anisotropic surface tension and kinetic attachment should be introduced \cite{takaki2014phase, steinbach2009phase} but have been neglected here for simplicity. The detailed derivation has been summarized in Appendix. The equation above is robust in propagating a diffuse interface using physical quantities such as kinetic coefficient $\mu$, Gibbs-Thomson coefficient $\Gamma$, interface width $\eta$, and undercooling $\Delta T$. The equation can be easily implemented in two or three dimensions without explicit front tracking. The value of the phase field $\phi$ indicates the local phase state. The solid and liquid phases are represented by 1 and 0, respectively. The interface shows a diffuse profile varying from 1 to 0. As the interface becomes diffuse, the concentration jump across the sharp interface in Fig.\ref{fig:models}a becomes a smooth transition in Fig.\ref{fig:models}b. 

\begin{figure}[ht]
    \centering
    \includegraphics[width=0.7\linewidth]{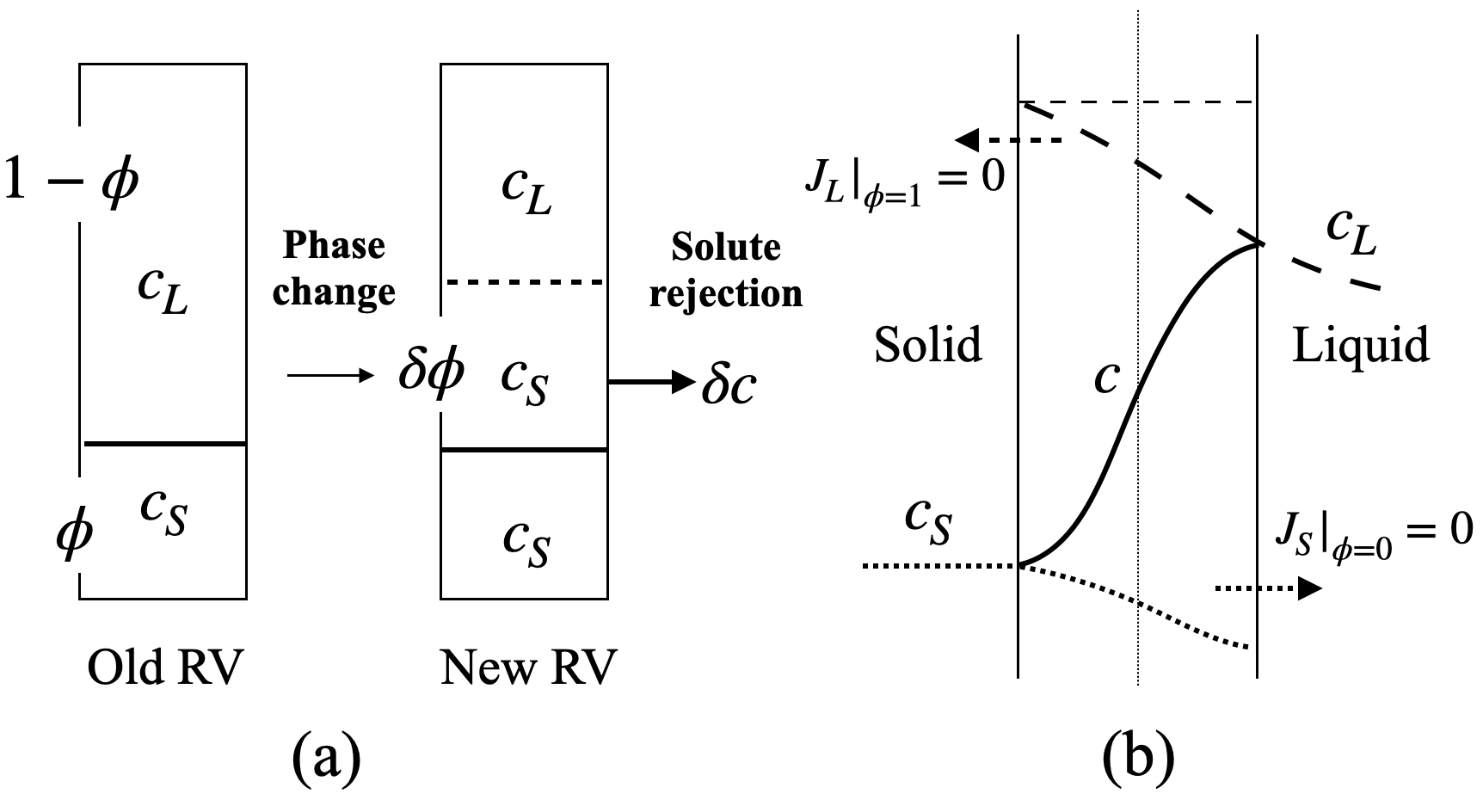}
    \caption{Solute partitioning in the standard alloy model. (a) Local change of the phase field followed by solute rejection in a  representative volume (RV) of the diffuse interface; (b) profiles of overall, solid, and liquid concentrations of a moving interface. The fluxes at the solid-interface and liquid-interface boundaries have been set to zero.}
    \label{fig:con3}
\end{figure}

The coupling of the diffusion equation into the phase-field equation is realized through the undercooling
\begin{equation}
\Delta T = T_{m} + m_{L}c^i_{L} -T_{i}, 
\label{eq:dT}
\end{equation}
\noindent which depends on the temperature $T_i$ and liquid concentration $c^i_{L}$ in the interface and is related to the melting temperature $T_m$ of solvent and liquidus slope $m_L$. The interface can be taken as a mixture of solid and liquid phases, of which the concentrations are calculated according to the relations of mass conservation and solute partitioning \cite{kim1999phase,beckermann1999, boettinger2002phase}, 
\begin{equation}
    c = \phi c_S + (1-\phi)c_L,
    c_S/c_L = k,
    \label{eq:cscl0}
\end{equation}
\begin{equation}
    c_L = \frac{c}{1+(k-1)\phi},
    \label{eq:cscl}
\end{equation}
\noindent in which $k$ can be the equilibrium partition coefficient determined from the phase diagram. As shown in Fig.\ref{fig:con3}a, the representative volume (RV) of the solid and liquid mixture is denoted by the phase fractions and concentrations. As a consequence of phase transition, the amount of rejected solute can be expressed by
\begin{equation}
\delta c = \delta \phi (c_{L}-c_{S}),
\label{eq:dlec}
\end{equation}
\noindent which could be simply released to the remaining liquid in the new RV
\begin{equation}
    c'_L = c_L + \frac{\delta c}{(1-\phi -\delta \phi)}.
    \label{eq:clp}
\end{equation}

\noindent The subsequent solute transport is realized by the Fick-like diffusion equation with separate fluxes in the two phases
\begin{equation}
     \frac{\partial c}{\partial t} = \nabla \big[ D_S\phi\nabla c_S + D_L(1-\phi)\nabla c_L \big ]
     \label{eq:dcdt}.
\end{equation}

The profiles of concentrations after phase transition and solute redistribution are illustrated in Fig.\ref{fig:con3}b. The diffusion layer in the bulk liquid extends into the diffuse interface. Thus, the interface undercooling has a gradient and inevitably includes the undercooling of the bulk liquid, which is not responsible for the interface motion \cite{steinbach2009phase}. One simple way to eliminate the bulk undercooling within the interface is to calculate using the liquid concentration at the solid-interface boundary, as shown by the dashed horizontal line in Fig.\ref{fig:con3}b and expressed by
\begin{equation}
     c^i_L |_{1>\phi>0} = c_L|_{\phi \rightarrow 1}
     \label{eq:coni}
\end{equation}
\noindent which can be implemented in numerical calculation by searching the grid near the solid-interface boundary along the interface normal $\vec{n} = \nabla \phi /|\nabla \phi|$. The flattening of the liquid concentration in the interface has profound effects that will be explained in the next section when the standard model above is extended to further approximate the sharp interface model. For now, the gradient of undercooling in the normal direction has been eliminated and interface velocity is expected to be lower than that of the sharp interface model because the concentration of $c_L|_{\phi \rightarrow 1}$ is spuriously high as the rejected solute can not be completely transported out of the artificially wide diffuse interface. Only when the interface width is very small, the simulation result can approximate the analytically true solution. 
 
\subsection{Middle-obstacle model}

\begin{figure}[ht]
    \centering
    \includegraphics[width=0.7\linewidth]{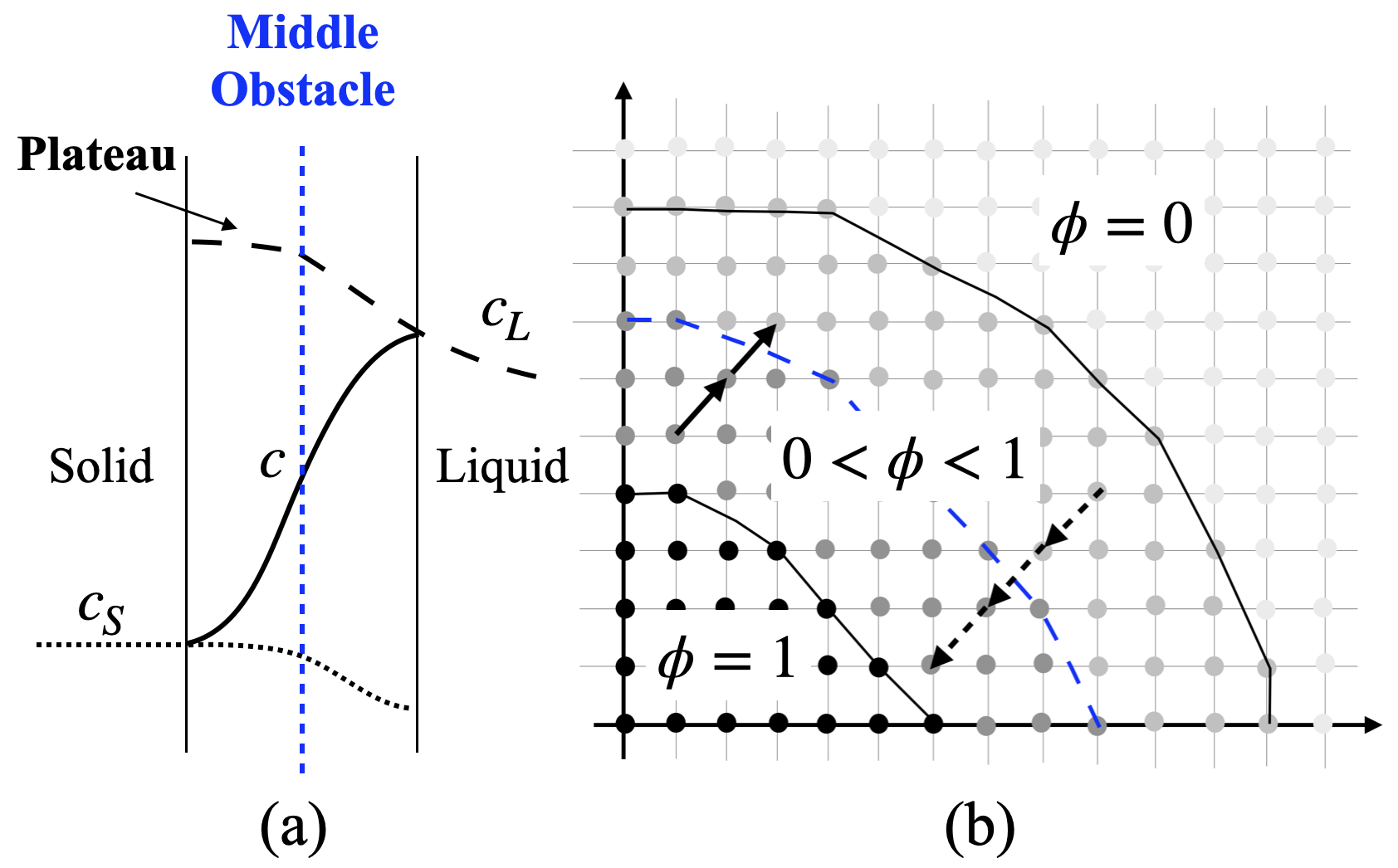}
    \caption{Emulating the sharp interface model through the middle-obstacle approach. (a) The concentration profiles after the implementation of the cast operation on the rejected solute behind the middle obstacle; (b) grid-searching within the diffuse interface along the interface normal vector.}
    \label{fig:mo}
\end{figure}

Intuitively, rather than releasing all the rejected solute $\delta c$ to the local liquid, $\delta c$ behind the middle plane of the diffuse interface can be cast forward to emulate the sharp interface model. The middle plane is called the ``middle obstacle'' for two reasons: it is in alignment with the double obstacle free energy density \cite{steinbach2009phase}; it acts as an imaginary obstacle that divides the diffuse interface into solid- and liquid-like parts. Consequently, the concentration profile behind the middle obstacle becomes a plateau (Fig.\ref{fig:mo}a) because of the absence of local release of the rejected solute. As illustrated in Fig.\ref{fig:mo}b, the cast operation is realized by searching the next grid along the normal vector until the phase field has a value of 0.5 and then updating the concentrations ($c_L$ and $c$) ahead of the middle obstacle. During the subsequent diffusion, the cast solute may flow back and change the liquid concentration behind the middle obstacle. The concentration of $c_L|_{\phi \rightarrow 1}$ at the backmost position of the diffuse interface is most likely to remain unaffected and elegantly emulates the interface concentration on the liquid side in the sharp interface model (Fig.\ref{fig:models}a). As the results of the undercooling flattening (Eq.(\ref{eq:coni})) along the interface normal, the effect of interface stretching should be eliminated and the effect of arc length diffusion on the interface kinetics should be minimized. Through this middle-obstacle approach, the diffuse interface is expected to be ``sharpened'' for accurately solving free boundary problems in alloy solidification.

\subsection{Incorporation of non-equilibrium effects}

The greatest advancement of the middle-obstacle model is the easiness of incorporating non-equilibrium effects \cite{aziz1982model, aziz1994transition, trivedi1994dendritic}. In practice, as the partitioning coefficient becomes a function of interface velocity, the calculation of interface concentrations using Eq.(\ref{eq:cscl}) suffers from numerical instability and errors, for which the reason is still unknown. The remedy is to calculate the concentrations through a strategy of solute capturing \cite{zhu2023middle}. The newly formed solid captures a portion of the solute in its previous liquid. The solid concentration is updated without affecting the liquid concentration
\begin{equation}
    c'_S = \frac{\phi c_S + \delta \phi \lambda c_L}{\phi + \delta \phi}, 
\end{equation}
\noindent The amount of rejected solute is 
\begin{equation}
\delta c = \delta \phi (1-\lambda)c_L.
\label{eq:dlec2}
\end{equation}
\noindent The capture coefficient $\lambda$ is related to the partitioning coefficient by 
\begin{equation}
    \lambda = \frac{\phi}{\delta \phi}(k - \frac{c_S}{c_L} ) + k \approx k, 
    \label{eq:lamb}
\end{equation}
of which the first term with local concentrations has been truncated so that the computational stability can be restored. The capture coefficient should be velocity-dependent and can be formulated by referring to the continuous growth (CG) model. Here, an expression is given as
\begin{equation}
    \lambda(v) = \frac{k_e + v/(\chi v_D)}{1+v/(\chi v_D)}
    \label{eq:capv}
\end{equation}
\noindent in which $v_D$ is the diffusive velocity in the CG model and $\chi$ is the factor used to compensate the loss of accuracy due to the truncation in Eq.(\ref{eq:lamb}). By regulating the factor, the partition coefficient in the numerical results can approximate that in the CG model. The interface velocity $v$ is determined by referring to the local change of phase fraction,
\begin{equation}
   v = \frac{\eta}{\pi \sqrt{\phi(1-\phi)}}\frac{\partial \phi}{\partial t}
   \label{eq:int-v}
\end{equation}
\noindent The variable $m_{L}$ in Eq.(\ref{eq:dT}) should be replaced by   
\begin{equation}
    m^v_{L} = \frac{1-k_v+ \big [k_v + (1-k_v)\alpha \big] \ln (k_v/k_e)}{1-k_e}m^e_{L}, 
    \label{eq:mlv}
\end{equation}
\noindent in which $\alpha$ denotes the extent of solute drag and $k_v$ is the local non-equilibrium partition coefficient defined by the concentration ratio at the solid-interface boundary
\begin{equation}
    k_v= \frac{c_S|_{\phi\rightarrow 1}}{c_L|_{\phi\rightarrow 1}}.
    \label{eq:kv}
\end{equation}

\section{Results and discussions}

\subsection{Equilibrium conditions}

\begin{figure}[t]
    \centering
    \includegraphics[width=0.7\linewidth]{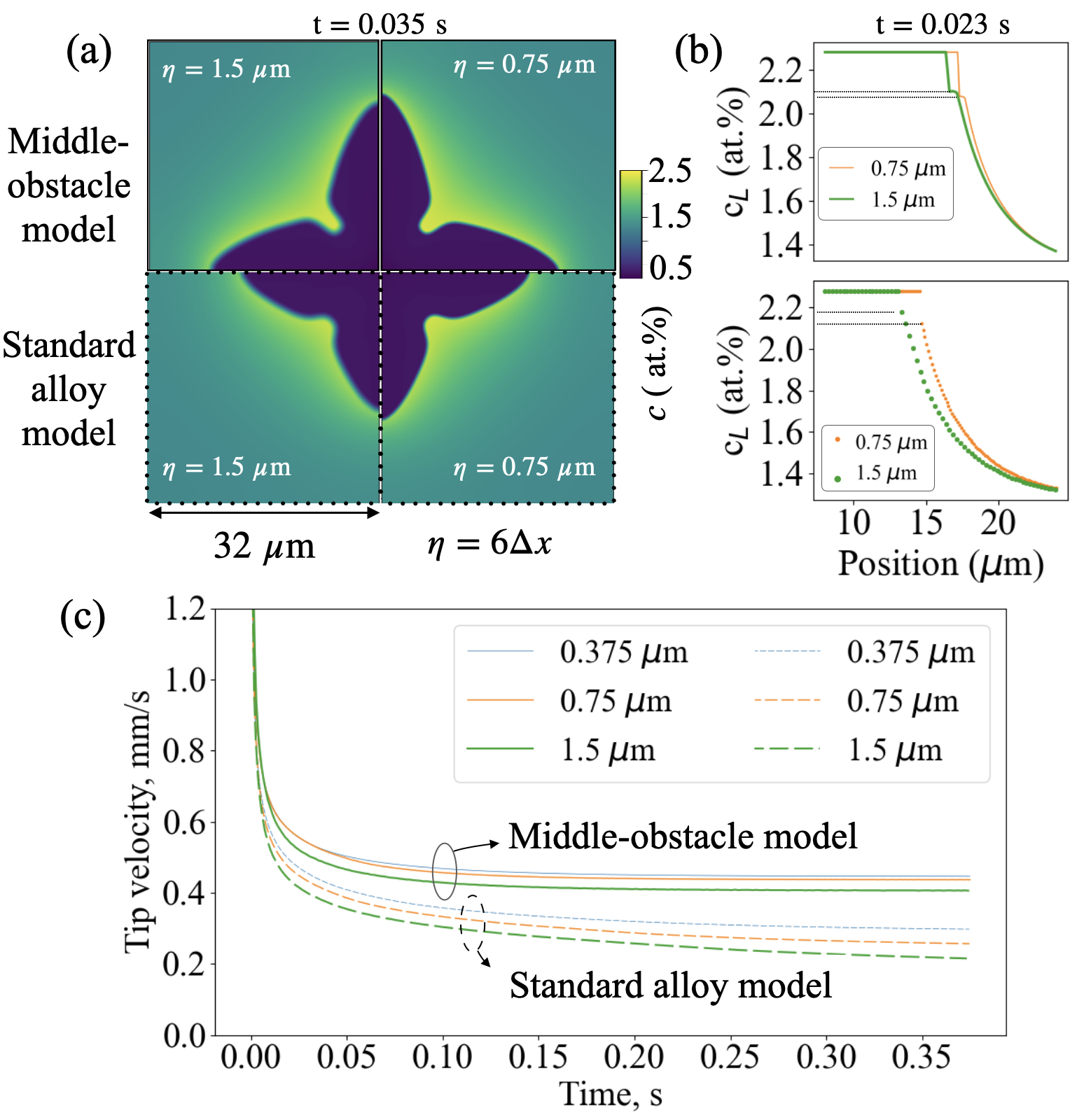}
    \caption{Simulation results of alloy solidification under equilibrium conditions for different interface widths. (a) Concentration maps during dendritic growth; (b) liquid concentration along the dendrite arms; (c) temporal variation of tip velocity in the moving frame for different interface widths. The middle-obstacle approach has dramatically reduced the width effect so that the invariant tip dynamics can be approached.}
    \label{fig:equi}
\end{figure}

The solidification simulation of Al-1.3at.(3wt.)\%Cu alloy under equilibrium and isothermal conditions has been conducted. The parameters are chosen as follows: $T_m$ = 931 K, $m_L$ = $-$600 K/at.\%, k = 0.14, $D_L$ = 2.4$\times 10^{-9}$ m$^2$/s, $D_S$ = $10^{-3}D_L$, $T_i$ = 917 K, $\Gamma$ = $2.4\times10^{-7}$ Km, $\mu$ = $5\times10^{-4}$ m/(sK). The strength coefficients of the cubic anisotropic function are 0.02  for surface tension and 0.1 for kinetic attachment. The interface width $\eta$  is about hundreds of times larger than the physical size and has been resolved by 6 grids. The results of the dendrite morphologies in the middle-obstacle model (MOM) and standard alloy model (SAM) are compared in Fig.\ref{fig:equi}a. The tip velocity in the MOM is higher and less sensitive to the change of interface width compared to those in SAM. The profiles of liquid concentrations along the dendrite arms are shown in Fig.\ref{fig:equi}b. They have the equilibrium value $c_{L}^e$ in the bulk solid and show the diffusion layer in the bulk liquid. Within the diffuse interface, instead of the continuous concentration gradient in the SAM, the concentration plateau appears in the MOM. The concentration of $c_L|_{\phi \rightarrow 1}$ in MOM is more resistant to the change of interface width, as indicated by the horizontal dashed lines.

The temporal changes of the tip velocities are shown in Fig.\ref{fig:equi}c. The computation frames in Fig.\ref{fig:equi}a moved with the dendrite tips for enough time until the steady states might be reached. The tip velocity in SAM increased slowly with the decreasing interface width and could hardly reach a steady state. By contrast, all the tips in MOM had reached the steady-state velocities, which tended to converge to an invariant value as the interface width becomes smaller. The convergence behavior of the MOM suggests that the middle-obstacle approach is indeed helpful for gaining accurate results with relatively large interface width and low grid resolution. Notably, the mathematical formulation and numerical implementation of the current model are extremely simple compared to previous quantitative phase-field models \cite{karma2001phase, carre2013implementation}, which take the asymptotic analysis for approximating the sharp interface model and are mostly one-sided models without consideration of solid diffusivity. In the middle-obstacle model, the two-sided diffusion is an inherent feature as expressed by Eq.(\ref{eq:cscl}-\ref{eq:dcdt}).

\subsection{Non-equilibrium conditions}

\begin{figure}
    \centering
    \includegraphics[width=0.7\linewidth]{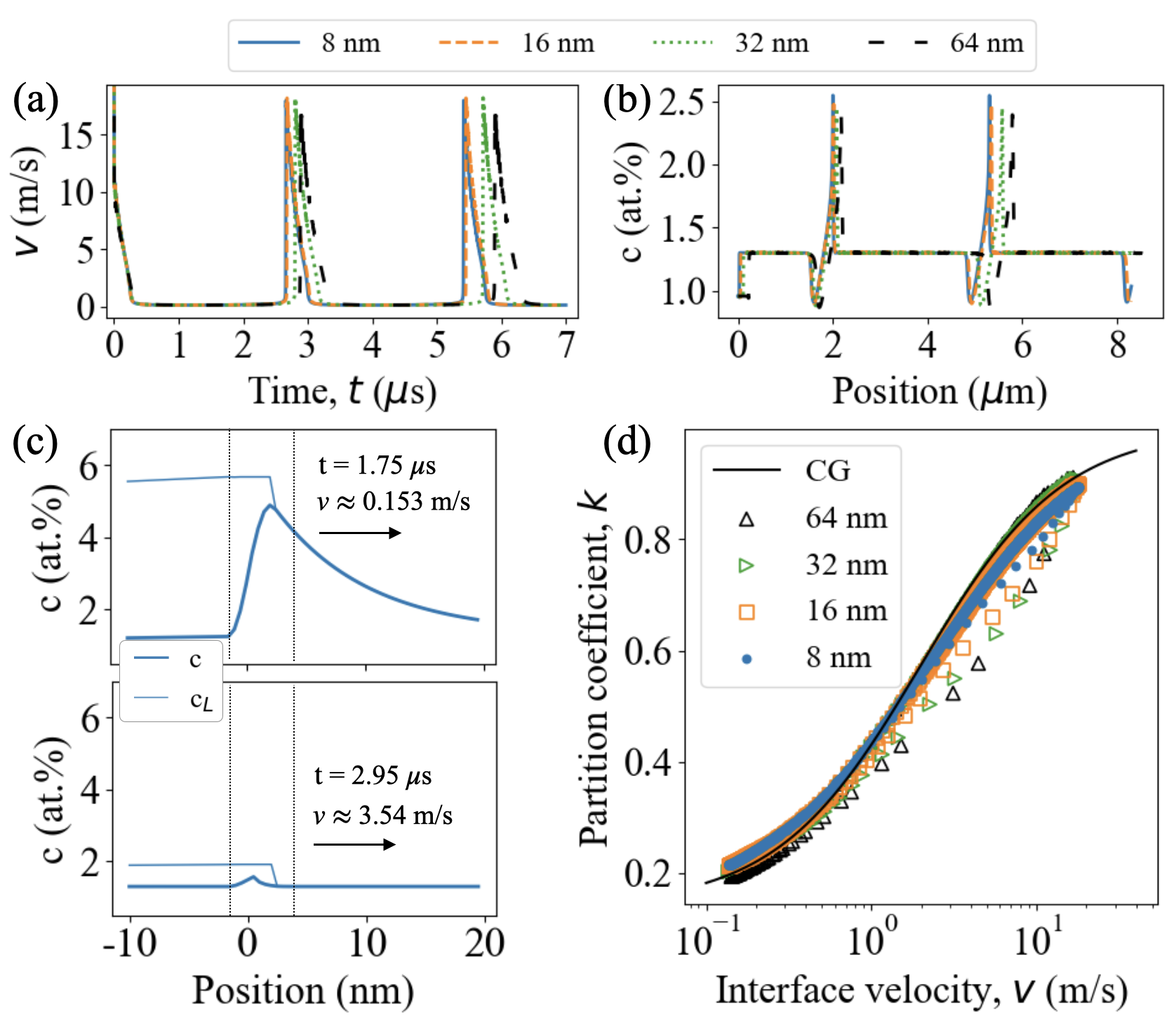}
    \caption{ One-dimensional (1D) non-equilibrium interface kinetics during rapid solidification. The banding phenomenon is reflected by oscillatory patterns of (a) temporal variation of the interface velocity and (b) concentration profile after solidification. (c) Segregated and non-segregated concentration profiles of interfaces with low and high velocities; (d) partitioning coefficients measured during the banding for different interface widths.}
    \label{fig:non-equi1}
\end{figure}

The simulation of rapid solidification of Al-1.3at.\%Cu alloy is conducted under the temperature gradient $G$ of 1.0$\times$10$^7$ K/m and cooling rate R of 1.2$\times$10$^7$ K/s. The material parameters are the same as those in the previous equilibrium calculation except for the kinetic coefficient, which can have a more realistic value of 0.5 m/(sK) because of the usage of nanometer interface width. The diffusive velocity $v_D$ is 1.96 m/s and the drag coefficient $\alpha$ is 0.645 \cite{ji2023microstructural}. The diffuse interface is resolved by 8 grids. As the one-dimensional (1D) interface is pulled by the moving temperature gradient, Fig.\ref{fig:non-equi1}a shows the interface velocity oscillates in the range from about 0.1 to 15.0 m/s, exhibiting the banding phenomenon when the interface velocity approaches the absolute stability \cite{trivedi1994dendritic}. The resulting concentration distribution after solidification (Fig.\ref{fig:non-equi1}b) shows similar patterns of oscillation. For different interface widths, the slightly shifted profiles suggest the anticipated resistance to the effects of the artificial interface width owing to the adoption of the middle-obstacle approach in the present model. The concentrations across the moving interface can have either segregated or non-segregated profiles, as shown in Fig.\ref{fig:non-equi1}c. The partition coefficient of the moving interface during banding oscillation has been monitored and compared with the analytical function of the CG model. The factor $\chi$ for the simulations with increasing interface widths is set to 0.8, 0.75, 0.7, and 0.65. As shown in Fig.\ref{fig:non-equi1}d, almost all the simulation data stay close to the analytical function. The very few deviated data points are caused by the deformed diffuse interface during the surge of interface velocity when the measurement of the concentrations at the solid-interface boundary becomes inaccurate. For other data points, the discrepancy with the CG model stays below 10\% regardless of the interface width. The accuracy can be further diminished by choosing a more suitable factor $\chi$ or developing a technique that can use the full expression of Eq.(\ref{eq:lamb}) as the input, which is open for future exploration. The present results of non-equilibrium interface kinetics are considerably accurate compared to the recent phase-field model \cite{ji2023microstructural}.

\begin{figure}
    \centering
    \includegraphics[width=0.7\linewidth]{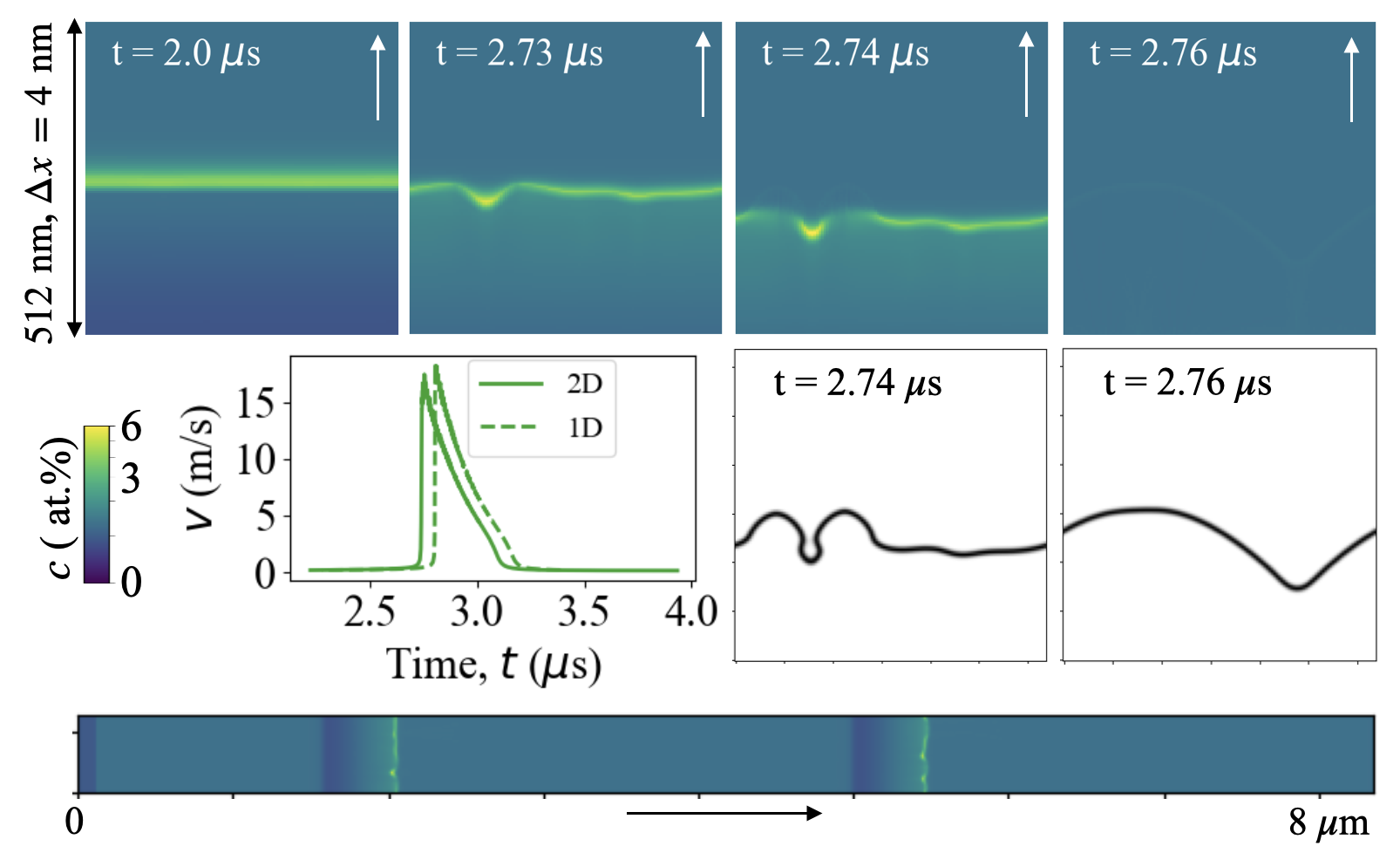}
    \caption{Burgeoning growth of non-equilibrium interface in two-dimensional (2D) space. First row: concentration maps of the growth front in a moving frame; second row: temporal change of front velocity and interface contours; third row: the resulting concentration distribution with banded pattern. }
    \label{fig:non-equi2}
\end{figure}

Finally, the burgeoning growth during the surge of interface velocity in two-dimensional space is showcased to further verify the applicability of the middle obstacle methodology. The physical and cooling conditions are the same as the previous 1D calculation. As shown in Fig.\ref{fig:non-equi2}, the solute accumulated ahead of the interface during the low-speed propagation, and the cellular morphology, namely Mullins-Sekerka instability, developed due to constitutional undercooling in front of the solid. The non-segregated growth was initiated at the cellular tips and expanded to the whole growth front (see the interface contours below the concentration maps). The solute segregation becomes almost invisible because the solid and liquid concentrations are the same and only a small concentration hump appears ahead of the interface. The front velocity is plotted against the one in the 1D calculation. The difference is caused by the 2D interface instability. The resulting concentration distribution in the bottom shows the oscillatory pattern and resembles the banded structure observed in the microstructure after rapid solidification \cite{gremaud1991banding}. 

\section{Summary and perspectives}

As one of the most discussed free boundary problems, alloy solidification under universal cooling conditions has been described through a unified diffuse and sharp interface model. The middle-obstacle methodology features in simplicity, accuracy, and seamless correlation with the sharp interface model with experimentally measured physical quantities. It deserves the attention of the microstructure modeling community and stands as an exemplar for addressing free boundary problems in the phenomena of phase transformation. With the help of this high-fidelity model, the development of solidification science and engineering, especially in the field of additive manufacturing \cite{koizumi2022digital, tourret2023morphological}, can be further promoted in the decades to come.

\section*{Acknowledgement}
The author extend the gratitude for the support received from the JSPS KAKENHI (Grant Number 22J11558).

\section*{Appendix: Derivation of the phase-field equations}

We start with a free energy functional $F$ that describes the total free energy of the system with the diffuse interface represented by
\begin{equation}
F = \int_V f(\phi, \nabla \phi) dv,
\tag{A1}
\end{equation}

\noindent in which $f$ is the free energy density for a given state of phase field $\phi$ and its gradient $\nabla \phi$. It can be written by the following expression with physical quantities of the phases and interface
\begin{equation}
f = \frac{4\gamma\eta}{\pi^2} a^2_s(\vec{n})|\nabla \phi|^2 + \frac{4\gamma}{\eta}\phi(1-\phi) + (1-h(\phi))G_0 + h(\phi)G_1,
\tag{A2}
\end{equation}

\noindent which includes interface energy $\gamma$, interace width $\eta$ and bulk free energy densities $G_{0/1}$. The function $a_s$ is dependent on the interface normal $\vec{n} = \nabla \phi/|\nabla \phi|$ and is used to incorporate anisotropic surface energy. The function $h(\phi)$ is used to interpolate the bulk free energy densities within the interface. The profile of the phase field that minimizes the total free energy can be found through the Euler-Lagrange equation for the free energy density
\begin{equation}
    \frac{\partial f}{\partial \phi} - \nabla \frac{\partial f}{\partial \nabla \phi} = 0.
\tag{A3}
\end{equation}

\noindent When the above equation is non-zero, the evolution of the phase field can be expressed by the following dissipative equation, which is also called the Allen-Cahn equation
\begin{equation}
\begin{aligned}
\tau \frac{\partial \phi}{\partial t} &=  \nabla \frac{\partial f}{\partial \nabla \phi} - \frac{\partial f}{\partial \phi} \\
& = \frac{8\gamma\eta}{\pi^2} \nabla \frac{\partial (\frac{a_s^2(\vec{n})}{2}|\nabla \phi|^2)}{\partial \nabla \phi} + \frac{8\gamma}{\eta}(\phi - \frac{1}{2}) + h'(\phi)(G_1-G_0).
\end{aligned}
\tag{A4}
\end{equation}

\noindent For isotropic surface energy, the orientation-dependent function $a_s(\vec{n})$ is unity and the governing equation is expressed by
\begin{equation}
    \tau \frac{\partial \phi}{\partial t} = \frac{8\gamma\eta}{\pi^2}\nabla^2\phi + \frac{8\gamma}{\eta}(\phi - \frac{1}{2}) + h'(\phi)(G_1-G_0).
\tag{A5}
\end{equation}

\noindent For stationary interface in 1D, the free energy difference is zero and the equation becomes
\begin{equation}
    \frac{8\gamma\eta}{\pi^2}\nabla^2\phi + \frac{8\gamma}{\eta}(\phi - \frac{1}{2}) = 0.
\tag{A6}
\end{equation}

\noindent The traveling wave solution that satisfies the above equation is expressed by
\begin{equation}
    \phi(x, t) = \frac{1}{2}-\frac{1}{2}\sin\frac{\pi}{\eta}x,  -\frac{\eta}{2}\leq x \leq \frac{\eta}{2}.
\tag{A7}
\end{equation}

\noindent As illustrated in Fig.A1, the geometrical relation between the spatial and temporal changes gives the following relation

\begin{equation}
\frac{\partial \phi}{\partial t} = -v \frac{\partial \phi}{\partial x}. \tag{A8}
\end{equation}

\noindent The gradient of this solution can be related to itself
\begin{equation}
    |\nabla \phi| = \frac{\pi}{\eta} \sqrt{\phi(1-\phi)}.
\tag{A9}
\end{equation}

\noindent With this relation, the excessive energy due to the diffuse interface can be integrated to test the validity of the original free energy functional
\begin{equation}
\begin{aligned}
F
&=\int \bigg [ \frac{4\gamma\eta}{\pi^2}|\nabla \phi|^2 + \frac{4\gamma}{\eta}\phi(1-\phi)\bigg ] dx \\
&=\frac{4\gamma}{\eta} \int \phi(1-\phi) dx  \\
&= \frac{\gamma}{2\eta}\int^{\eta/2}_{-\eta/2} [1+\cos(\frac{2\pi}{\eta}x)]dx \\
&= \gamma .
\end{aligned} \tag{A10}
\label{eq:gamma}
\end{equation}

\noindent For the moving interface, it can be assumed that the diffuse profile is identical to the stationary one. The governing equation is reduced to be
\begin{equation}
    \tau \frac{\partial \phi}{\partial t} =  h'(\phi)(G_1-G_0).
\tag{A11}
\end{equation}

\noindent By substituting the interface velocity to the left-hand side
\begin{equation}
  \tau |\nabla \phi|v = h'(\phi)(G_1-G_0).
\tag{A12}
\end{equation}

\noindent If the variable of the phase field on the two sides of the equation can be canceled, the equation above is reduced to the reaction-rate equation by which the interface velocity is linearly proportional to the free energy difference when surface tension is not considered for the planar interface
\begin{equation}
    v = K\Delta G.
\tag{A13}
\end{equation}

\noindent As mentioned before, the function $h(\phi)$ is used for weighing the free energy densities in the diffuse interface. It has been found that the following form is suitable for canceling $|\nabla \phi|$
\begin{equation}
h(\phi) = \frac{1}{\pi}\big [(4\phi-2)\sqrt{\phi(1-\phi)} + \arcsin(2\phi-1) \big] + \frac{1}{2}.
\tag{A14}
\label{eq:hphi}
\end{equation}

\noindent of which the derivation to $\phi$ is 
\begin{equation}
    h'(\phi) = \frac{8}{\pi}\sqrt{\phi(1-\phi)}.
\tag{A15}
\end{equation}

\noindent Consequently, the dissipative parameter $\tau$ can be related to the kinetic coefficient
\begin{equation}
    \tau = \frac{8\eta}{\pi^2K}.
\tag{A16}
\end{equation}

\noindent Finally, the phase-field equation with physical parameters is expressed by
\begin{equation}
    \frac{\partial \phi}{\partial t} =  K \bigg \{ \gamma \bigg[ \nabla^2\phi +\frac{\pi^2}{\eta^2} \bigg(\phi-\frac{1}{2}\bigg)\bigg]+ \frac{\pi\sqrt{\phi(1-\phi)}}{\eta}\Delta G \bigg \}, 
\tag{A17}
\end{equation}

\noindent which can have another form with a temperature difference
\begin{equation}
\boxed{
    \frac{\partial \phi}{\partial t} =  \mu \bigg \{ \Gamma \bigg[ \nabla^2\phi +\frac{\pi^2}{\eta^2} \bigg(\phi-\frac{1}{2}\bigg)\bigg]+ \frac{\pi\sqrt{\phi(1-\phi)}}{\eta}\Delta T \bigg \}
}.
\tag{A18}
\end{equation}

\noindent For interface with anisotropic surface tension and attachment kinetics, the governing equation with the orientation-dependent functions is 
\begin{equation}
    \frac{\partial \phi}{\partial t} =  \mu a_k(\vec{n}) \bigg \{ \Gamma \bigg[ \nabla \frac{\partial (\frac{a^2_s(\vec{n})}{2}|\nabla \phi|^2)}{\partial \nabla \phi} +\frac{\pi^2}{\eta^2} \bigg(\phi-\frac{1}{2}\bigg)\bigg]+ \frac{\pi\sqrt{\phi(1-\phi)}}{\eta}\Delta T \bigg \}.
\tag{A19}
\end{equation}

\setcounter{figure}{0}
\renewcommand{\thefigure}{A\arabic{figure}}
\begin{figure}[h]
\centering
\includegraphics[width=0.8\textwidth,keepaspectratio]{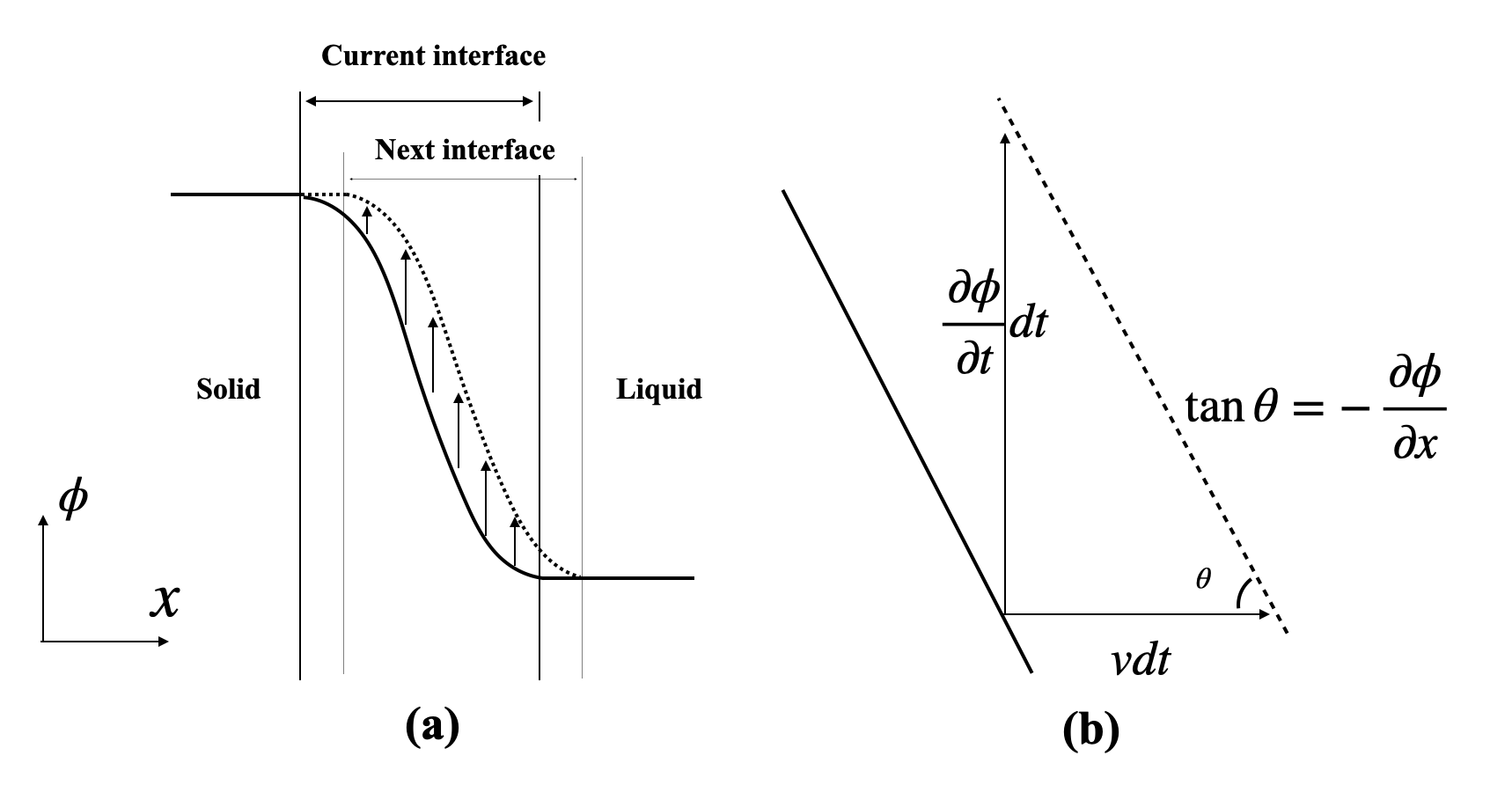}
\caption{Derive the velocity of the interface from the local changes of the phase field.} 
\label{fig:vdp}
\end{figure}

\end{document}